\documentclass[aps,prl,twocolumn,showpacs,floatfix,preprintnumbers]{revtex4}
\usepackage{graphicx}
\usepackage{amssymb}
\usepackage{epstopdf}
\usepackage{dcolumn}
\usepackage{bm}
\usepackage{amsmath}

\begin{document} 
\preprint{FERMILAB-PUB-07-555-A-CD-E-TD}
\title{Search for axion-like particles using a variable baseline photon regeneration technique}

\author{A.~S. Chou$^{1,2}$}
\author{W. Wester$^1$}
\author{A. Baumbaugh$^1$}
\author{H.~R. Gustafson$^3$}
\author{Y. Irizarry-Valle$^1$}
\author{P.~O. Mazur$^1$}
\author{J.~H. Steffen$^1$}
\author{R. Tomlin$^1$}
\author{X. Yang$^1$}
\author{J. Yoo$^1$}
\affiliation{$^1$Fermi National Accelerator Laboratory, PO Box 500, Batavia, IL 60510\\
$^2$Center for Cosmology and Particle Physics, New York University, 4 Washington Place, New York, NY 10003\\
$^3$Department of Physics, University of Michigan, 450 Church St, Ann Arbor, MI 48109}

\date{\today}
\begin{abstract}
We report the first results of the GammeV experiment, a search for milli-eV mass particles with axion-like couplings to two photons.  The search is performed using a ``light shining through a wall'' technique where incident photons oscillate into new weakly interacting particles that are able to pass through the wall and subsequently regenerate back into detectable photons.  The oscillation baseline of the apparatus is variable, thus allowing probes of different values of particle mass.  We find no excess of events above background and are able to constrain the two-photon couplings of possible new scalar (pseudoscalar) particles to be less than $3.1\times 10^{-7} \mbox{ GeV}^{-1}$ ($3.5\times 10^{-7} \mbox{ GeV}^{-1}$) in the limit of massless particles.

\end{abstract}
\maketitle
Recently, the PVLAS experiment reported a positive signal in a photon oscillation experiment \cite {Zavattini:2005tm}.  In their ``disappearance'' experiment, a polarized 1064 nm laser beam was sent into an Fabry-Perot cavity which also contained a high transverse magnetic field.  By measuring a net rotation of the laser polarization vector upon exiting the cavity, PVLAS found a non-zero relative attenuation of laser polarization components transverse to and parallel to the external magnetic field.  This result was interpreted as evidence for the production of new spin-0 particles  from one of the polarization components \cite{Raffelt:1987im}.  Newer PVLAS results reported at conferences \cite{gastaldi} included the measurements of non-zero polarization rotation with 532 nm green light, and non-zero ellipticity at both wavelengths, where the relative phase delay is induced by loop effects. 

The effective axion-like interaction Lagrangian for a pseudoscalar particle is:
\begin{equation}
\label{E:pseudoscalar}
{\mathcal L}_{\rm{pseudoscalar}}
  =-\frac{g}{4}\phi F_{\mu\nu}\widetilde{F}^{\mu\nu}
  =g \phi(\vec{E}\cdot\vec{B})
\end{equation}
while that for a scalar particle is:
\begin{equation}
\label{E:scalar}
{\mathcal L}_{\rm{scalar}}
  =-\frac{g}{4}\phi F_{\mu\nu}{F}^{\mu\nu}
  = g \phi (\vec{E}\cdot\vec{E}-\vec{B}\cdot\vec{B}).
\end{equation}
While PVLAS was not able to determine the parity of the new particle, their four measurements gave a consistent picture of a low scalar mass $m_\phi\sim 1.2\mbox{ meV}$ and a rather large two-photon coupling $g \sim 2.5\times 10^{-6} \mbox{ GeV}^{-1}$.   The suggested PVLAS region of interest was not previously excluded by the pioneering BFRT laser oscillation experiment \cite{Cameron:1993mr} \cite{Ruoso:1992nx}.  However, for reasons unknown, the PVLAS signal disappeared after the experimental apparatus was later rebuilt in an effort to improve the detection \cite{Zavattini:2007ee}.  A number of experimental efforts world-wide have begun to test the hypothesis that the anomalous PVLAS observations are due to a new axion-like particle \cite{Patras}, and in particular the BMV experiment has recently excluded the pseudoscalar interpretation of the PVLAS signal\cite{Robilliard:2007bq}\footnote{OSQAR has also recently reported results \cite{Pugnat:2007nu} which may not be valid due to a matching of group velocities but not phase velocities of photon and scalar waves in their residual gas technique.}.    

New low-mass particles with masses smaller than the electron mass have yet to be directly detected, but models containing such light degrees of freedom have been motivated by the small neutrino mass differences \cite{Fardon:2003eh} \cite{Kaplan:2004dq}, by the vacuum dark energy density of $(2 \mbox{ meV})^4$ \cite{Brax:2007ak}, and rather generically by string theory \cite{Svrcek:2006yi} and theories with large extra dimensions \cite{Maity:2007un}.  Couplings of light particles to photons are tightly constrained by star cooling considerations, including the experimental limits set by the CAST axion helioscope \cite{Andriamonje:2007ew} whose recent limit $g < 8.8\times 10^{-11} \mbox{ GeV}^{-1}$ strongly excludes PVLAS.  Various methods have been postulated, however, to evade the star cooling limits  \cite{Brax:2007ak}\cite{Jaeckel:2006xm}\cite{Ganguly:2006ki}.

In this letter we report the first results from \mbox{GammeV}, a photon regeneration experiment similar to that originally suggested in \cite{VanBibber:1987rq}.  Our apparatus is specifically designed to quickly probe the region in parameter space suggested by the PVLAS results, at modest expense.  As an ``appearance'' experiment rather than a ``disappearance'' experiment, a positive signal would yield an unambigious new interpretation of oscillations of photons into new, weakly interacting particles.  

\section{Experimental design}
\begin{figure}[t]
\includegraphics[width=0.48\textwidth]{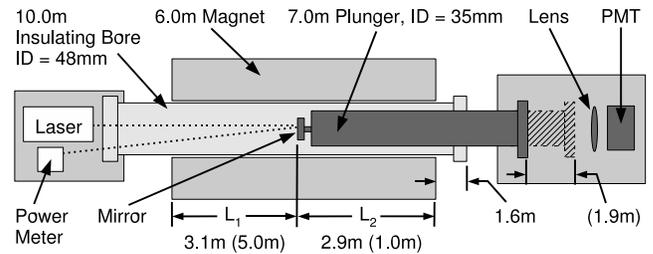}
\caption{\label{F:apparatus} Diagram (not to scale) of the experimental apparatus.  The initial vacuum chamber consists of a 10 m insulating warm bore which is offset by 1.6 m from the end of the 6 m magnetic field region, and is sealed to the sliding plunger via a double o-ring assembly.  The sliding plunger has a range of motion of 1.9 m, and contains an independent vacuum chamber.  The vacuum window at the far end slides within a stationary, long dark box.}
\end{figure}

The key to this experiment is the short 5 ns, 160 mJ pulses of 532 nm light emitted with a repetition rate of 20 Hz by our light source, a frequency-doubled Continuum Surelite I-20 Nd:YAG laser.  As described below, the small duty cycle enables a large reduction in the detector noise via coincidence counting.  The laser light is vertically polarized and when needed, a halfwave plate is used to obtain horizontal polarization.  The laser pulses are sent through a vacuum system (diagrammed in Fig.~\ref{F:apparatus}) designed around an insulating warm bore inserted into a 6 m Tevatron superconducting dipole magnet.  The magnet produces a 5 T vertical field uniform across the aperture of the $48 \mbox{ mm}$  inner diameter warm bore.  A ``wall'' consisting of a high-power laser mirror on the end of a long (7 m) hollow stainless steel ``plunger'' is inserted into the warm bore.   The mirror may be placed at various positions within the magnet by sliding the plunger.  The plunger mirror projects the reflected wave into a photon state and the transmitted wave into a (pseudo-)scalar state, provided that the scalar is sufficiently weakly-interacting to pass through the material of the mirror.  The mirror also functions to reflect the incident laser power out of the magnet to prevent heating of the magnet coils.  The mirror is mounted on a welded stainless steel cap on the end of the plunger in order to prevent stray photons from passing through.  Thus, the beam passing through the end of the plunger is a pure scalar beam.  These scalars can then oscillate back into photons through the remaining magnetic field region inside the $35 \mbox{ mm}$ inner diameter hollow plunger.  Upon exiting the magnetic field region, the interaction ceases and the photon-scalar wavefunction is frozen.  This combined wavefunction then propagates $\sim 6\mbox{ m}$ into a dark box where a Hamamatsu H7422P-40 photomultiplier tube (PMT) module is used to detect single photons in coincidence with the laser pulses.  As described below, a high signal to noise ratio is achieved due to the very short pulses emitted by the laser.

The photon-scalar transition probability may be written in convenient units as:
\begin{eqnarray}
\label{E:conversionprob1}
P_{\gamma \rightarrow \phi} &=& \frac{4 B^2\omega^2}{M^2 (\Delta m^2)^2} \sin^2 \left( \frac{\Delta m^2 L}{4\omega} \right) \\
\label{E:conversionprob2}
&\approx&  \frac{4 B^2\omega^2}{M^2 m_\phi^4} \sin^2 \left( \frac{m_\phi^2 L}{4\omega} \right) \\
 \label{E:conversionprob3}
&=& 1.5\times 10^{-11} \frac{(B/\mbox{Tesla})^2(\omega/\mbox{eV})^2}{(M/10^5 \mbox{ GeV})^2 (m_\phi/10^{-3} \mbox{ eV})^4} \nonumber  \\
& & \times \sin^2 \left( 1.267 \frac{(m_\phi/10^{-3} \mbox{ eV})^2 (L/\mbox{m})}{(\omega/\mbox{eV})} \right)
\end{eqnarray}
where $B$ is the strength of the external magnetic field, $\omega$ is the inital photon energy, $L$ is the magnetic oscillation baseline.   The mass-squared difference between the scalar mass and the effective photon mass, $\Delta m^2=m_\phi^2-m_\gamma^2$, characterizes the mismatch of the phase velocities of the photon wave and the massive scalar wave and determines the characteristic oscillation length.  While the photon does not really gain a mass in a normal dielectric medium, the phase advance may be modelled with an effective imaginary mass 
$m_\gamma^2 = -2 \omega^2 (n-1)$,
where $n$ is the index of refraction \cite{vanBibber:1988ge}.  Both the warm bore and the interior of the plunger are pumped to moderate vacuum pressures of less than $10^{-4} \mbox{ Torr}$, and a conservative estimate gives $\sqrt{-m_\gamma^2} <  10^{-4} \mbox{ eV}$.  Therefore, starting with Eqn.~\ref{E:conversionprob2} we assume that the contribution from the effective photon mass is negligible for larger values of $m_\phi$ near the PVLAS region.

As can be seen from Eqn.~\ref{E:conversionprob3}, the meter scale baseline provided by typical accelerator magnets is well-suited for probing the milli-eV range of possible particle masses.  This fact can be a curse as well as a boon because for a monochromatic laser beam, a fixed magnet length may accidentally coincide with a minimum in the oscillation rather than a maximum.  Indeed, this is a possible reason why the BFRT experiment \cite{Ruoso:1992nx} did not see the PVLAS signal even though they had similar sensitivity.  GammeV's plunger design allows us to change the oscillation baseline and thus scan through all possible values of the scalar mass in the milli-eV range without any regions of diminished sensitivity.  The total conversion and regeneration probability contains two factors of Eqn.~\ref{E:conversionprob2}, corresponding to the pre-mirror and post-mirror magnetic field regions of lengths $L_1$ and $L_2$.  The total probability varies as $\sin^2 \left(\frac {m_\phi^2 L_1}{4\omega} \right) \sin^2 \left(\frac{m_\phi^2 L_2}{4\omega} \right)$ where $L_1+L_2=6\mbox{ m}$.

To detect regenerated photons we use a $51\mbox{ mm}$ diameter lens to focus the beam onto the $5 \mbox{ mm}$ diameter GaAsP photocathode of the PMT.  The alignment is performed using a low power green helium-neon alignment laser and a mock target.  The alignment is verified both before and after each data-taking period by replacing the sealed plunger with an open-ended plunger, re-establishing the vacuum, and firing the Nd:YAG laser onto a flash paper target.  An optical transport efficiency of $92\%$ is measured using the ratio of laser power transmitted through the open-ended plunger and through the various optics and vacuum windows, to the initial laser power, using the same power meter in both cases to remove systematic effects.  The quantum efficiency of the photocathode is factory-measured to be $38.7\%$ while the collection efficiency of the metal package PMT is estimated to be $70\%$.  The PMT pulses are amplified by 46 dB and then sent into a NIM discriminator.  Using a highly attenuated LED flasher as a single photon source, the discriminator threshold is optimized to give 99.4$\%$ efficiency for triggering on single photo-electron pulses while also efficiently rejecting the lower amplitude noise.   By studying the trigger time distribution, the deadtime fraction due to possible multiple rapid PMT pulses is found to be negligible (0.001$\%$).  Thus, we estimate the total photon transport and counting efficiency to be $(25\pm 3)\%$.  Using this threshold, and the built-in cooler to cool the photocathode to $0^\circ$C, we measure a typical dark count rate of 130 Hz. 

\begin{figure}[t]
\includegraphics[width=0.48\textwidth]{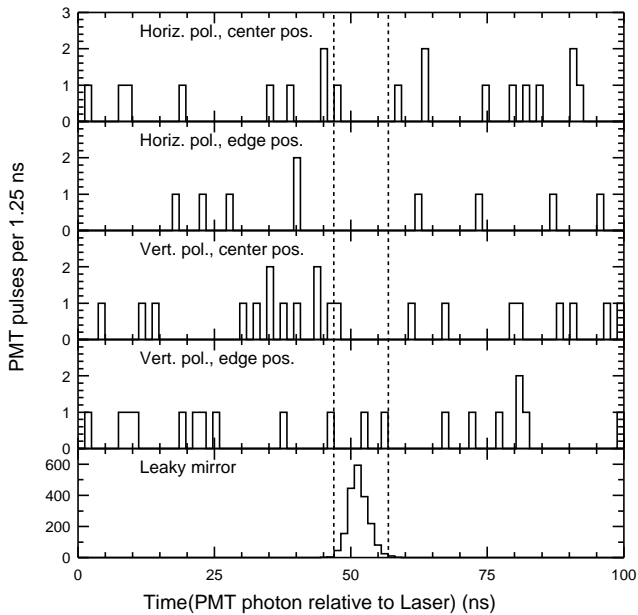}
\caption{\label{F:timing} PMT trigger times for the four run configurations, shown relative to the expected time distribution of photons as calibrated from the leaky mirror data.}
\end{figure}

To perform the coincidence counting we use two Quarknet boards \cite{quarknet} \cite{quarknetgps} with 1.25 ns timing precision, referenced to a GPS clock.  The Quarknet boards determine the absolute time of the leading edge of time-over-threshold triggers from the PMT and from a monitoring photodiode that is located inside the laser box.  The clocks on the laser board and on the PMT board are synchronized using an external trigger from a signal generator.

The absolute timing between the laser pulses and the PMT traces is established by removing the plunger with the mirror, and allowing the laser to shine on the PMT through several attenuation stages consisting of two partially reflective (``leaky'') mirrors, a pinhole, and multiple absorptive filters mounted directly on the aperture of the PMT module.  The $10^{19}$ photons per second emitted by the laser are thus attenuated to a corresponding PMT trigger rate of less than 0.1 Hz for this timing calibration and to provide an {\it in situ} test of the data acquisition system.

The regenerated photons should arrive at the same time as the straight-through photons since milli-eV particles are also highly relativistic.  For coincidence counting between the laser pulses and the PMT, a 10 ns wide window is chosen and includes $99\%$ of the measured photon time distribution shown in Fig.~\ref{F:timing}.  The coincident dark count rate can be estimated to be $R_{\rm{noise}}=20\mbox{ Hz}\times 130\mbox{ Hz}\times 10\mbox{ ns} = 2.6\times 10^{-5} \mbox{ Hz}$.  This noise rate is negligible to the expected signal rate of $\sim 2\times 10^{-3} \mbox{ Hz}$ estimated from the central values of the PVLAS parameters.  
\begin{table}
\begin{tabular}{|l|c|c|c|c|}
\hline
Configuration&$\#$ photons&Est.Bkgd&Candidates&g[GeV$^{-1}$]\\
\hline
Horiz.,center&$6.3\times 10^{23}$&$1.6$&1&$3.4\times 10^{-7}$\\
Horiz.,edge&$6.4\times 10^{23}$&$1.7$&0&$4.0\times 10^{-7}$\\
Vert.,center&$6.6\times 10^{23}$&$1.6$&1&$3.3\times 10^{-7}$\\
Vert.,edge&$7.1\times 10^{23}$&$1.5$&2&$4.8\times 10^{-7}$\\
\hline
\end{tabular}
\caption{\label{F:table} Summary of data in each of the 4 configurations.}
\end{table}

\section{Data collection}
To cover the entire PVLAS range of $m_\phi$, $20 \mbox { hours}$ of data are collected in each of four configurations of plunger position and polarization.   The plunger is placed either in the center of the magnet with $L_1 = 3.1 \mbox{ m}$ and $L_2 = 2.9 \mbox{ m}$ or near the far edge of the magnet with $L_1 = 5.0 \mbox{ m}$ and $L_2 = 1.0 \mbox{ m}$.  With each plunger position, data are taken separately with both vertically and horizontally polarized laser light in order to test pseudoscalar and scalar couplings, respectively.  The central plunger position covers most of the PVLAS signal region but the corresponding oscillation probability goes to zero in Eqn.~\ref{E:conversionprob3} at $m_\phi\approx 1.4 \mbox{ meV}$ where the argument of the each of the sine factors approaches $\pi$.  These zeroes indicate regions of diminished sensitivity as the scalars have oscillated back into photons upon reaching the plunger mirror.  Moving the plunger to the far position simultaneously changes the baseline for both the initial oscillation and the regeneration, and shifts the two regions of diminished sensitivity away from $m_\phi=1.4 \mbox{ meV}$.  The two plunger positions thus cover the entire PVLAS signal region.

The operating conditions are continuosly monitored during each run.  The reflected beam from the plunger mirror is slightly offset from the incident beam, and is directed into a calorimetric power meter by a pick-off mirror.  The number of incident photons is determined with $3\%$ accuracy from these measurements.  The alignment of the laser is monitored using a fast solid state camera which takes 30 Hz of images of the reflected laser spot on the pick-off mirror.  The total pathlength to the pick-off mirror is comparable to the pathlength to the PMT, and so transverse deviations seen in the images are closely matched to the deviations at the PMT.  Small $\sim$mm scale transverse deviations are seen during the course of a typical 5 hour run, due to small changes in the orientation of the plunger mirror as the plunger slowly cools through heat leaks in the warm bore insulation.  The deviations are small enough that, were they due to actual changes in the laser alignment, the beam would still clear the aperture of the plunger.  In addition, the focussing lens at the PMT makes the final light collection system insensitive to these potential sub-mrad angular deviations.  Nevertheless, the alignment is double-checked using the open-ended plunger after collecting data in each configuration, and no misalignment has ever exceeded our tolerances.  The operation of the PMT is monitored using its dark rate.  In addition, an LED flasher fires every 5 minutes to verify the integrity of the light collection system.

\begin{figure}[t]
\includegraphics[width=0.48\textwidth]{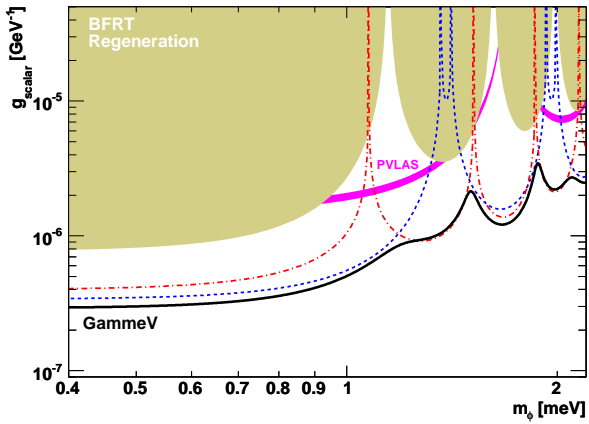}
\caption{\label{F:scalar} 3$\sigma$ limit contours for scalar particles.  The solid black line is the combined limit using data taken at both the central (red dot-dashed) and the edge (blue dashed) plunger positions.  The PVLAS rotation signal (pink/dark grey) and the BFRT regeneration limit (tan/light grey) are also shown.}
\end{figure}

\begin{figure}[t]
\includegraphics[width=0.48\textwidth]{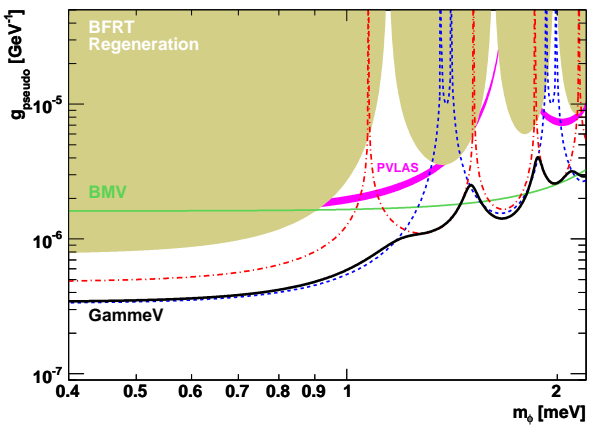}
\caption{\label{F:pseudoscalar} 3$\sigma$ limit contours for pseudoscalar particles.  }
\end{figure}

\section{Results}
We count the PMT triggers within the 10 ns coincidence time window, defined {\it a priori} by the leaky mirror data.  The expected background is measured using the dark counts in time bins outside of the coincidence window.  No excess counts above background are observed in any of the four configurations.  The data are summarized in Tab.~\ref{F:table}.  The PMT timing data, along with the leaky-mirror calibration data, are shown in an expanded time scale in Fig.~\ref{F:timing}.  No excess counts above background are observed in any time bins near the laser pulse.  For the central values of the PVLAS parameters one would expect $\sim$150 excess counts.

We use the Rolke-Lopez method \cite{Rolke:2004mj} to obtain limits on the regeneration probabilities, and use Eqn.~\ref{E:conversionprob2} to obtain the corresponding $3\sigma$ upper bounds on the coupling $g$ as a function of $m_\phi$.  The relative systematic uncertainties of $12\%$ on the photon transport and detection, and $3\%$ on the laser power measurement are incorporated in the limits.  The GammeV limits are shown in Figs.~\ref{F:scalar} and \ref{F:pseudoscalar} along with the PVLAS 3$\sigma$ signal region, the BFRT 3$\sigma$ regeneration limits, and the $99.9\%$ limit on pseudoscalar couplings from BMV.  As expected, the regions of insensitivity for one plunger position are well-covered by using the other plunger position.  Data from both plunger positions are combined and analyzed jointly to produce the combined limit curve.  The weakly-interacting axion-like particle interpretation of the PVLAS data is excluded at more than $5\sigma$ by GammeV data for both scalar and pseudoscalar particles.  The asymptotic $3\sigma$ upper bounds on $g$ for small $m_\phi$ for each configuration are listed in Tab.~\ref{F:table}, and the combined analysis gives  $3.1\times 10^{-7} \mbox{ GeV}^{-1}$ ($3.5\times 10^{-7} \mbox{ GeV}^{-1}$) for the scalar (pseudoscalar) couplings.    The GammeV exclusion region extends beyond the previous best limits and sets limits in regions where BFRT had reduced sensitivity.

{\it Acknowledgements:} We would like to thank the staff of the Fermilab Magnet Test Facility for their tireless efforts, and the technical staff of the Fermilab Particle Physics Division design group who aided in the design and construction of the apparatus.  JS thanks the Brinson Foundation for their generous support.  This work is supported by the U.S. Department of Energy under contract No. DE-AC02-07CH11359.  AC is also supported by NSF-PHY-0401232.

\end{document}